\documentclass[11pt]{article}
\usepackage{amsmath,amsthm,amssymb,graphicx}
\usepackage[letterpaper,margin=1in]{geometry}

\newcommand{\GG}{\Gamma}
\newcommand{\Gp}{\psi}
\newcommand{\En}{{\cal{E}}}

\newcommand{\Ga}{\alpha}

\newcommand{\Gd}{\delta}

\newcommand{\Gf}{\phi}
\newcommand{\Gg}{\gamma}
\newcommand{\Gk}{\kappa}
\newcommand{\Gl}{\lambda}
\newcommand{\GL}{\Lambda}
\newcommand{\Gm}{\mu}

\newcommand{\Gs}{\sigma}

\newcommand{\Gth}{\theta}

\newcommand{\dd}{{\hbox{\rm d}}}

\newcommand{\Eq}[1]{Eq.~(\ref{eq:#1})}

\newcommand{\prt}{\partial}
\newcommand{\povr}[2]{\frac{\prt #1}{\prt #2}}

\newcommand{\Figure}[1]{Figure~\ref{fig:#1}}

\usepackage{sectsty}
\subsectionfont{\it}

\usepackage{apacite}

\title{Measurement scale in maximum entropy models of species abundance}

\author{\\ Steven A.\ Frank\footnote{Department of Ecology and Evolutionary Biology, University of California, Irvine, CA 92697--2525, USA, email: safrank@uci.edu}\\}

\begin{document}

%% apacite modifications
%% special renew commands to get style close to JEB formatting

\renewcommand\BOthers{\emph{et al}\hbox{}}
\renewcommand\BOthersPeriod{\emph{et al}\hbox{}}
\renewcommand{\BCBT}{}				% ref list no final comma w/two authors 
\renewcommand{\BCBL}{}				% ref list no final comma w/more authors
\renewcommand{\APACrefYearMonthDay}[3]{#1}	% no parens around year in reflist
\renewcommand{\APACrefYear}[1]{#1}	% no parens around year in reflist
\renewcommand{\BEd}{edn\hbox{}}         	% edition
\renewcommand{\APACrefYear}[1]{%
  {#1}%
}
\renewcommand{\APACjournalVolNumPages}[4]{%
  \Bem{#1}%             journal
  \ifx\@empty#2\@empty
  \else
    \unskip\ \textbf{#2:}%  volume
  \fi
  \ifx\@empty#3\@empty
  \else
    \unskip%      issue number
  \fi
  \ifx\@empty#4\@empty
  \else
    \unskip\ {#4}%      pages
  \fi
}
\renewcommand{\APACaddressPublisher}[2]{%
  \ifx\@empty#1\@empty
    \ifx\@empty#2\@empty
    \else
      {#1}%                 address
    \fi
  \else
    {#2}%                   publisher
    \ifx\@empty#2\@empty
    \else
      \unskip, {#1}%        address
    \fi
  \fi
}
		% must put after begin doc

\maketitle

\vskip1in
%\noindent\textit{Running head:} Measurement scale and pattern\hfill\break
%\noindent\textit{Article type:} Ideas and Perspectives\hfill\break
%\noindent\textit{Abstract words:} 178\hfill\break
%\noindent\textit{Manuscript words:} 9700\hfill\break
%\noindent\textit{Main text words:} 7500\hfill\break
%\noindent\textit{Number of references:} 43\hfill\break
%\noindent\textit{Number of figures:} 1\hfill\break
%\noindent\textit{Number of tables:} 0\hfill\break
%\noindent\textit{Keywords:}  Macroecology, maximum entropy, measurement theory, neutral theory, symmetry \hfill\break

%\linenumbers

%\newpage
\begin{abstract}

The consistency of the species abundance distribution across diverse communities has attracted widespread attention.  In this paper, I argue that the consistency of pattern arises because diverse ecological mechanisms share a common symmetry with regard to measurement scale.  By symmetry, I mean that different ecological processes preserve the same measure of information and lose all other information in the aggregation of various perturbations.  I frame these explanations of symmetry, measurement, and aggregation in terms of a recently developed extension to the theory of maximum entropy.  I show that the natural measurement scale for the species abundance distribution is log-linear: the information in observations at small population sizes scales logarithmically and, as population size increases, the scaling of information grades from logarithmic to linear.  Such log-linear scaling leads naturally to a gamma distribution for species abundance, which matches well with the observed patterns.  Much of the variation between samples can be explained by the magnitude at which the measurement scale grades from logarithmic to linear.  This measurement approach can be applied to the similar problem of allelic diversity in population genetics and to a wide variety of other patterns in biology.  

%\vskip1in
%
%\centering 
%
%{\bf Keywords:} maximum entropy
%
%\vskip0.5in
%
%{\bf Running head:} Symmetry and probability

%\vfill\setcounter{page}{2}\thispagestyle{plain}

\end{abstract}
\newpage

%% If using separate title page, page numbers get messed up
%\setcounter{page}{3}

\begin{quote}
\textit{It is better to be vaguely right than exactly wrong \cite{read09logic}.}
\end{quote}

\section*{Introduction}

The species abundance distribution (SAD) describes the number of individuals of each species observed in a sample.  The SAD shape is remarkably consistent across communities.  Many distinct rare species each have only a single individual in the sample. The number of different species declines as the count of individuals per species rises \shortcite{fisher43the-relation,preston48the-commonness,macarthur57on-the-relative,macarthur60on-the-relative,whittaker65dominance,may75patterns,hubbell01the-unified,magurran04measuring,mcgill07species,ulrich10a-meta-analysis}.  

Such a consistent pattern naturally leads to widespread interest.  What is the best description of the pattern?  What theory best explains the consistency across such diverse habitats?  How should we think of the differences in distribution that do occur between certain types of habitat? A vast literature is devoted to these questions.  Here, I focus on connecting two points of view in the recent debate.  

The first point of view argues that different processes can lead to the same pattern.  Thus, a fit between the observed SAD and a particular mechanistic theory must be treated with caution, because other equally plausible mechanisms lead to the same pattern \shortcite{may75patterns,pueyo07the-maximum,mcgill10towards}.  No one argues directly against this point of view.  Nonetheless, the tendency of different processes to lead to the same pattern is sometimes ignored, because we do not have a fully convincing theory for why widely different processes would in fact lead consistently to the narrow range of observed SAD patterns.

The second point of view uses maximum entropy theory to explain the observed SAD. In maximum entropy, the most likely probability distribution is the one that is most random, or has highest entropy, subject to certain minimal constraints.  A constraint might, for example, be that the average population size of species is set by the habitat.  By maximum entropy, the abundances of species would be the most random pattern such that the overall average abundance is fixed \shortcite{shipley06from,pueyo07the-maximum,harte08maximum,banavar10applications,haegeman10entropy,he10maximum}.

Maximum entropy could, in principle, explain why different mechanistic hypotheses lead to the same SAD pattern.  Two different mechanisms have the same SAD shape if they both constrain the average abundance and otherwise produce various perturbations that ultimately tend to cancel in the aggregate. However, there is at present no general understanding of the relation between maximum entropy and the tendency for different mechanistic hypotheses to converge to the same SAD pattern.  Thus, maximum entropy is often viewed as an alternative theory for SAD patterns \cite{mcgill10towards}, rather than a more fundamental principle about probability that necessarily plays a central role in translating process into pattern.

In this paper, I use recent advances in maximum entropy theory to strengthen the argument that many different underlying mechanistic hypotheses lead to the same common SAD pattern.  The advances follow from my work with Eric Smith, showing the importance of information invariance and measurement scale in understanding the fundamental ways in which different probability patterns arise \cite{frank10measurement,frank11a-simple}.

The first section describes the common theoretical SADs that have been used to fit observed patterns.  In that section, I show that the gamma distribution subsumes the log series, power law, and geometric distributions as special cases.  The gamma distribution can also take on shapes very close to the widely used lognormal distribution.  The gamma often fits observed SAD patterns better than the lognormal. 

The second section argues that the gamma distribution arises as the natural expression of pattern on a log-linear measurement scale.  A log-linear scale is logarithmic at small magnitudes and continuously grades into a linear scale at large magnitudes.  In terms of SADs, the match to the gamma pattern means that the information one obtains from an observed species abundance in a sample scales logarithmically at low abundance and linearly at high abundance. I use recent advances in maximum entropy theory to derive this relation between log-linear measurement scale and observed SAD patterns \cite{frank10measurement,frank11a-simple}. 

From the first two sections, I conclude that SADs often follow a gamma distribution, and the gamma distribution arises naturally as the expression of pattern on a log-linear measurement scale.  Those conclusions leave us with the question:  Why do ecological mechanisms often lead to log-linear scaling?  My main goal is to establish that question, which the first two sections accomplish.  In the third section, I explore possible answers by examining the way in which specific ecological mechanisms associate with log-linear scaling.  

The discussion analyzes the position of maximum entropy among the various approaches to understanding biological pattern.  \citeA{mcgill10towards} recently classed maximum entropy as an approach that makes particular hidden assumptions about mechanism. By this view, maximum entropy is a testable hypothesis that can be evaluated by observation.  By contrast, I argue that maximum entropy is like the calculus.  One does not test the calculus by comparing predictions with data.  Rather, both the calculus and maximum entropy provide analytical tools that help in understanding the logical relations between assumptions and observations.  

In the appendix, I note that \citeA{pueyo07the-maximum} originally established the basic approach of maximum entropy and invariance for species abundance problems.  I then describe specific limitations in the way that \citeA{pueyo07the-maximum} framed the maximum entropy problem and how my measurement theory approach resolves those problems.  My resolution connects the maximum entropy method to a broader framework of measurement and information, providing a deeper understanding that is essential for interpreting ecological pattern.

With regard to ecological pattern, maximum entropy can be used in two ways.  First, from a consistently observed pattern, such as the SAD, one can induce the necessary and sufficient attributes that various ecological mechanisms must have to match observed pattern.  Second, one can deduce the predicted pattern generated by a wide class of ecological mechanisms that share common attributes.  Those shared attributes determine the measurement scale and thus define pattern.  I will emphasize that the way in which maximum entropy has been applied to ecology needs to be revised to relate the method to the structure of the biological problem.

Common measurements of genetic diversity are often analogous to the species abundance problem.  In genetics, one may classify genetic variants into distinct alleles and then measure the abundance of each allele.  The distribution of allelic abundances in a sample has the same structure as the distribution of species in a sample.  The equivalence of species and allelic sampling problems has been discussed often \shortcite[Chapter 41]{watterson74models,hubbell01the-unified,leigh07neutral,johnson97discrete}.  In this paper, I focus on the ecological problem of species, because that subject has developed more fully the particular issues that I will analyze.  

Maximum entropy methods in relation to ecological and genetic patterns illustrate a deeper problem in biology \cite{frank09the-common}. How do we separate commonly observed patterns generated by typical processes of measurement and aggregation from those special patterns that provide information about the underlying biological mechanisms?  Current biological analysis has largely ignored this central problem.  Consistent progress depends on clearer understanding.  In particular, we must have some sense of what is surprising and informative versus what is unsurprising and uninformative.  Otherwise, much analysis devotes attention to what is in fact expected based on the simplest notions of aggregation and measurement.

\section*{SADs follow the gamma distribution}

Among $S$ species, the probability $p_y$ is the fraction of species each with $y$ individuals, and $Sp_y$ is the number of species each with $y$ individuals. The distribution of $p_y$ defines the SAD.  I focus on the underlying distribution of species abundances, ignoring the inevitable fluctuations caused by sampling.  Sampling fluctuations are important, but in this paper I wish to isolate the forces that shape the underlying distribution from sampling fluctuations.

In this section, I argue that a gamma probability distribution is a simple and general description of observed SAD patterns.  \shortciteA{ulrich10a-meta-analysis} recently conducted a meta-analysis of 558 SADs derived from 306 publications.  Their conclusions match the broad consensus in the literature that SADs typically fit best to either a log series distribution or a lognormal distribution.  They also found that several observed SADs fit best to a power law distribution.  

One can always quibble about the methods of fitting and the choice of alternative distributions to compare.  My point of view in this paper does not depend on the fine points of fitting alternative distributions. Rather, we can simply take the qualitative conclusion that observed distributions usually fit reasonably well to either the log series, lognormal, or power law pattern.  Different observed SADs vary in which of these three distributions fits best.

\subsection*{Log series and power law distributions}

The log series distribution has the form
\begin{equation}\label{eq:logseries}
  p_y = k\frac{\Gth^y}{y},
\end{equation}
where $k=-1/\log(1-\Gth)$, $0<\Gth<1$, and $y=1,2,\ldots$ is the number of individuals in the sample for each species of class $y$.  I use ``$\log$'' for the natural logarithm.  This distribution has its highest value (mode) at $y=1$, so that the rarest species, each represented by a single individual, occur most frequently in the sample. Species with increasingly large populations in the sample occur at a decreasing frequency.

The power law distribution has the form
\begin{equation}\label{eq:powerlaw}
  p_y = k y^{-\Gg},
\end{equation}
where $\Gg>1$, and $k$ is chosen so that the total probability is one over all values of $y$.  The power law also has a mode at one, and frequency declines steadily for species with increasing population sizes.  The rate of decline in frequency with increasing population size is slower for the power law than for the log series.

For discrete distributions, such as the forms of the log series and power law given here, we may truncate the distribution so that the greatest value of $y$ is not higher than some upper bound, $B$, resetting $k$ so that the total probability is one.

\subsection*{Normal and lognormal distributions}

To describe SADs by the lognormal, one must first transform the underlying measurement scale.  The traditional approach follows the Preston plot method, in which one forms bins on a $\log_2$ scale for abundance values $y$, such that $y=1$ maps to the $2^0$ or $\log_2=0$ bin, $y=2$ maps to the $\log_2=1$ bin, the combination of $y=3,4$ maps to the $\log_2=2$ bin, the combination of $y=5,6,7,8$ maps to the $\log_2=3$ bin, and so on.  This binning creates a discrete distribution on the logarithmic scale.  

On the logarithmic scale, the lognormal has the symmetric shape of a normal distribution.  For fits to observed SADs, one matches a continuous normal distribution to the discrete binned logarithmic distribution.

In the next section, I will describe the gamma distribution on both the linear and logarithmic scales.  To compare those forms of the gamma with the lognormal, it is useful to describe the transformations betweeen the log and linear scales for the lognormal. 

Here, I use $y$ for the linear scale and $x=\log y$ for the log scale.  I use the standard notation for continuous probability, in which $p_x\dd x$ is the probability that $x$ falls in the interval between $x$ and $x+\dd x$ for a small increment $\dd x$.  The magnitude of the increment $\dd x$ may change in ways that define the measurement scale, as shown in the following.

On the log scale, the normal distribution is
\begin{equation}\label{eq:lognormal}
  p_x\,\dd x = k e^{-\frac{(x-\Gm)^2}{2\Gs^2}}\dd x,
\end{equation}
where the measure $\dd x$ is for the log scale, $x$. To get the linear scale measure, we make the substitution $x=\log y$ to obtain
\begin{equation*}
  p_{\log y}\,\dd\log y = k e^{-\frac{(\log y-\Gm)^2}{2\Gs^2}}\dd \log y,
\end{equation*}
where $k$ is always taken to adjust so that the total probability is one.  We get the form of $p_y$ on the linear scale by changing the measure $\dd\log y=\dd y/y$ and noting that $p_y=p_{\log y}/y$, yielding
\begin{equation*}
  p_y\,\dd y = k y^{-1} e^{-\frac{(\log y-\Gm)^2}{2\Gs^2}}\dd y.
\end{equation*}
Thus, the distribution here is normal on the log scale and lognormal on the linear scale.  

To compare the lognormal and gamma distributions with log-binned SAD data on the logarithmic scale, we need to express the gamma on the logarithmic scale.  

\subsection*{Gamma and exponential-gamma distributions}

In this section, I first describe the gamma distribution on the linear scale and then present what I call the exponential-gamma on the log scale.

The discrete gamma distribution has the form
\begin{equation}\label{eq:discreteGammaE}
  p_y = k y^{\Ga-1}e^{-\Gl y}
\end{equation}
for $y=1,2,\ldots$.  Define $\Gth=e^{-\Gl}$, which allows us to write the discrete gamma as
\begin{equation}\label{eq:discreteGamma}
  p_y = k y^{\Ga-1}\Gth^y.
\end{equation}
The discrete gamma contains the discrete log series, power law, and geometric distributions as special cases.  For $\Ga=0$, we obtain the log series distribution; for $\Gth=1$, we obtain the power law distribution; for $\Ga=1$, we obtain the geometric distribution.  The gamma distribution subsumes the other classic distributions because it derives from a generalization of the measurement scales of the other distributions, as explained later.  

Because the two-parameter discrete gamma distribution contains the one-parameter log series, power law, and geometric distributions as special cases, the gamma must always fit any observed SAD at least as well as the special cases.  

In fitting observed distributions, it is widely known that the gamma and lognormal distributions can take on similar shapes and are often hard to distinguish in practice \shortcite{cho04a-comparison,silva07analysis}.  For SADs, the main weakness of the lognormal is that its symmetry on the log scale usually underestimates the frequency of rare species in the lower tail of the distribution \cite{diserud00a-general,hubbell01the-unified,wilson04biodiversity}.  A greater weight in the lower tail of the gamma is exactly the main difference between the gamma and lognormal distributions.  I show an example below.

To examine the gamma distribution on the log scale, we first need the continuous form on the linear scale. The continuous linear form has the same expression as the discrete form but with a continuous interpretation
\begin{equation*}
  p_y\,\dd y = k y^{\Ga-1}e^{-\Gl y}\dd y,
\end{equation*}
where $y>0$ and $k=\Gl^\Ga/\GG(\Ga)$, and I write $\dd y$ to emphasize the continuous measure of probability on the linear scale.  To transform this linear scale version to the log scale, we will once again use the substitution $x=\log y$.  However, in this case we are going in the reverse direction, so we need the inverse substitution $y=e^x$ yielding
\begin{align*}
  p_{e^x}\,\dd e^x &= k (e^x)^{\Ga-1}e^{-\Gl e^x}\dd e^x\\
          &= k e^{(\Ga-1)x - \Gl e^x}\dd e^x.
\end{align*}
Now change the measure to the log scale in terms of $x$ by the substitution $\dd e^x = e^x\dd x$ and note that $p_x=p_{e^x}e^x$, yielding what I call the exponential-gamma distribution
\begin{equation}\label{eq:expGamma}
  p_x\,\dd x = k e^{\Ga x - \Gl e^x}\dd x.
\end{equation}
Because $x=\log y$, and $y>0$, the domain is $-\infty<x<\infty$.  However, it is often useful to study the truncated form with $x\ge0$, corresponding to $y\ge1$ for species abundance counts that have a lower bound of one. In the truncated form, $k=\Gl^\Ga/\GG(\Ga,\Gl)$, where the denominator is the incomplete gamma function evaluated from a lower bound of $\Gl$.

In the case of the lognormal, that distribution is normal on the logarithmic scale and lognormal on the linear scale.  In the case here, the distribution is gamma on the linear scale and exponential-gamma on the logarithmic scale.

\subsection*{Example comparison of lognormal and gamma distributions}

\begin{figure}
\centering
\includegraphics[width=3.5in]{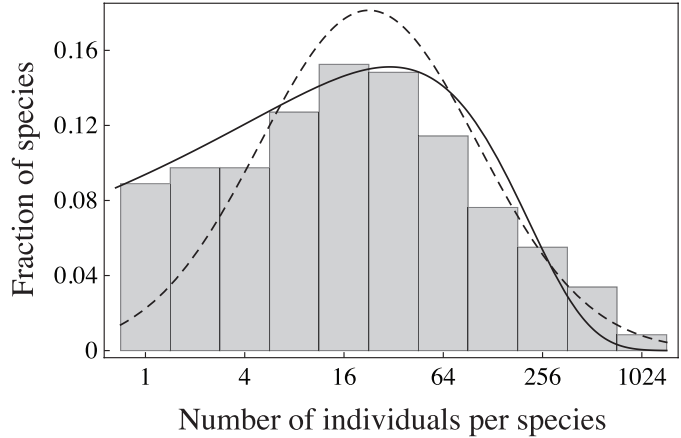}
\caption{Preston plot of species abundance distribution for tree species with diameter at breast height (dbh) greater than 10 cm from a 50 ha plot on Barro Colorado Island, Panama.  Data extracted from Figure 5.7 of \protect \citeA{hubbell01the-unified}. The numbers on the abcissa show the linear counts, $y$, scaled logarithmically so that $x=\log y$.  The dashed line shows a matching normal distribution on the log scale (lognormal on the linear scale) from \Eq{lognormal}.  The solid line shows a matching exponential-gamma distribution on the log scale (gamma on the linear scale) from \Eq{expGamma}.}\label{fig:bci1}
\end{figure}

In this paper, rather than fitting data to distributions, I emphasize the gamma distribution as a natural expression for the diversity of observed SAD forms.  However, it is useful to look at a plot of some data to get a feel for the shapes. \Figure{bci1} shows data for a typical SAD.  I fit by eye the matching normal and exponential-gamma distributions on the log scale, corresponding to the lognormal and gamma distributions on the linear scale.  

This figure illustrates the commonly observed excess of rare species compared with the lognormal pattern.  The gamma pattern differs most strongly from the lognormal by allowing a higher probability weighting of small values; otherwise the lognormal and gamma distributions are similar.  

\citeA{plotkin02sampling} commented on the good fit to SADs provided by the gamma distribution.  They also noted that the gamma distribution is not commonly used to fit SADs, in spite of the generally good match and some clear precedents \cite{fisher43the-relation,dennis84the-gamma,engen96population}. The gamma may lack popularity because explicit models of ecological process and sampling often lead to other distributions.  

In my view, those explicit models make the mistake of setting too many exact assumptions about the generation of pattern. Those exact assumptions can never be matched by the heterogeneous reality of nature.  Instead, the dominant aspects of pattern may arise from very general aspects of the way in which heterogeneous perturbations combine in the aggregate.  The central limit theorem is the most obvious example, in which linearly scaled perturbations lead to a normal distribution.  When perburbations follow other measurement scales, different distributions may arise.

Other distributions may sometimes provide a better fit than the gamma, for example, Hubbell's \citeyear{hubbell01the-unified} zero-sum multinomial distribution of the neutral theory. However, my point goes beyond the particular fit by various distributions with different numbers of parameters.  Rather, I emphasize that the gamma distribution subsumes as special cases several classic distributions commonly used for SADs, the gamma typically outperforms the lognormal in fitting, and the gamma arises naturally from very general aspects of measurement and information.  I am particularly interested in this last aspect of the gamma distribution as a simple expression of likely probability patterns in relation to natural changes in measurement scale with magnitude.

\section*{Gamma from log-linear measurement and maximum entropy}

In this section, I show that the gamma distribution arises as the natural expression of pattern on a log-linear measurement scale.  A log-linear scale is logarithmic at small magnitudes and continuously grades into linear at larger magnitudes. The following section illustrates why SADs may tend to associate with log-linear scaling, in which low population abundances carry information in relation to a log scale, and large population abundances carry information in relation to a linear scale.

To explain the claim that the gamma distribution expresses log-linear scaling, I first review the standard method of maximum entropy to derive probability distributions.  I then describe recent extensions to maximum entropy to incorporate the role of measurement scale.  My review of maximum entropy and description of the extensions for measurement scale are condensed summaries of \citeA{frank10measurement,frank11a-simple}.

\subsection*{Maximum entropy}

The method of maximum entropy defines the most likely probability distribution as the distribution that maximizes a measure of entropy (randomness) subject to various information constraints \cite{jaynes03probability}.  The idea is that the many random perturbations that affect pattern mostly tend to cancel each other in the aggregate, leaving the aggregate completely random except for any constraints that restrict the pattern.  

For example, the average number of individuals per species may be constrained by the productivity of the habitat, and that average will be maintained in spite of the wide variety of other processes that perturb species distributions.  So the final pattern must reflect that constraint.  Particular processes may tend to push species abundances in one direction, but other processes will push in the other direction.  Only the constraints remain in the aggregate, all else tends to maximum randomness or entropy.  Maximum randomness is equivalent to minimum information.  Thus maximizing entropy is equivalent to minimizing the information expressed in the final pattern subject to any constraints that cause information to be retained.

The power of maximum entropy is that aggregate patterns almost always seem to converge to a few simple patterns that express maximum randomness subject to just a few informational constraints.  To analyze a problem by maximum entropy, one first identifies the informational constraints that define a particular problem.  Then, by maximizing randomness subject to those constraints, one obtains the predicted form of the probability distribution that describes the pattern.  Going the other way, each common probability distribution is an exact expression of a few informational constraints with all else maximally random.

To derive a probability distribution by maximum entropy, we write the quantity to be maximized as
\begin{equation}\label{eq:maxEnt}
	\GL = \En - \Gk C_0 - \sum_{i=1}^n\Gl_iC_i,
\end{equation}
where $\En$ measures entropy, the $C_i$ are the constraints to be satisfied, and $\Gk$ and the $\Gl_i$ are the Lagrange multipliers to be found by satisfying the constraints.  Let $C_0=\sum p_y -1$ be the constraint that the probabilities must total one, where $p_y$ is the probability distribution function of $y$.  The other constraints are usually written as $C_i= \sum p_yf_i(y) -\bar{f}_i$, where the $f_i(y)$ are various transformed measurements of $y$, and the overbar denotes mean value. A mean value is either the average of some function applied to each of a sample of observed values, or an a priori assumption about the average value of some function with respect to a candidate set of probability laws. If $f_i(y)=y^i$, then $\bar{f}_i$ are the moments of the distribution---either the moments estimated from observations or a priori values of the moments set by assumption.  The moments are often regarded as ``standard'' constraints, although from a mathematical point of view, any properly formed constraint can be used.  

Here, I confine the analysis to a single constraint of measurement. I express that constraint with a more general notation, $C_1= \sum p_y T(f_y) -\bar{T}_f$, where $f_y\equiv f(y)$, and $T(f_y)\equiv T_f$ is a transformation of $f_y$.  I could, of course, express the constraining function for $y$ directly through $f_y$.  However, I wish to distinguish between an initial function $f_y$ that can be regarded as a standard measurement, in any sense in which one chooses to interpret the meaning of standard, and a transformation of standard measurements denoted by $T_f$ that arises from information about the measurement scale.  

The maximum entropy distribution is obtained by solving the set of equations
\begin{equation}\label{eq:maxEntSoln}
	\povr{\GL}{p_y} = \povr{\En}{p_y} - \Gk - 
	\Gl T_f=0,
\end{equation}
where one checks the candidate solution for a maximum and obtains $\Gk$ and $\Gl$ by satisfying the constraint on total probability and the constraint on $\bar{T}_f$. For continuous probability distributions, I assume that I can treat the entropy measures, the contraints, and the maximization procedure by the continuous limit of the discrete case.

In the standard approach, one defines entropy by extension of Shannon information 
\begin{equation}\label{eq:shannonDef}
	\En=-\int p_y\log\left(\frac{p_y}{m_y}\right)\dd y.
\end{equation}
For discrete distributions, $m_y$ is a prior probability distribution that sets the default pattern of randomness in the absence of any additional informational constraints, yielding an expression that can be interpreted as relative entropy, sometimes called the Kullback-Leibler divergence \cite{cover06elements}.  Alternatively, in the case of continuous distributions, $m_y$ may be an adjustment to maintain an invariant measure of information under changes in scale \cite{jaynes03probability,frank11a-simple}.  

With these definitions, the solution of \Eq{maxEntSoln} is
\begin{equation}\label{eq:shannonSolnMy}
	p_y \propto m_ye^{- \Gl T_f},
\end{equation}
where $\Gl$ satisfies the constraint $C_1$, and the proportionality is adjusted so that the total probability is one by choosing the parameter $\Gk$ to satisfy the constraint $C_0$.  In the applications in this paper, I will use $m_y\propto 1$ so that the default distribution is uniform, the most random pattern with the highest entropy, and the pattern that lacks any information.  I include in $T(f_y)$ any attributes of measurement that may deform the default distribution \cite{frank10measurement,frank11a-simple}.  The general maximum entropy solution with $m_y \propto 1$ is
\begin{equation}\label{eq:shannonSoln}
	p_y \propto e^{- \Gl T_f}.
\end{equation}

\subsection*{Information invariance and measurement scale}

Maximum entropy must capture all of the available information about a particular problem.  One form of information concerns transformations to the measurement scale that leave the most likely probability distribution unchanged \cite{jaynes03probability,frank09the-common,frank10measurement}.  Here, it is important to distinguish between measurements and measurement scale.  In my notation, I start with measurements, $f_y$, made on the measurement scale $y$.  For example, one may have measures of squared deviations about zero, $f_y=y^2$, with respect to the measurement scale $y$, such that $\bar{f}_y$ is the second moment of the measurements with respect to the underlying measurement scale.  

Suppose that one obtains the same information about the underlying probability distribution from measurements of $f_y$ or transformed measurements, $G(f_y)$.  Put another way, if one has access only to measurements $G(f_y)$, one has the same information that would be obtained if the measurements were reported as $f_y$.  One may say that the measurements $f_y$ and $G(f_y)$ are equivalent with respect to information, or that the transformation $f_y \rightarrow G(f_y)$ is an information invariance that describes a symmetry of the measurement scale.

To capture this information invariance in maximum entropy, we must express measurements so that 
\begin{equation}\label{eq:transDef}
  T(f_y) = \Gd + \Gf T[G(f_y)]
\end{equation}
for some arbitrary constants $\Gd$ and $\Gf$.  Putting this definition of $T(f_y)\equiv T_f$ into \Eq{shannonSoln} shows that the same maximum entropy solution arises from the observations $f_y$ or the transformed observations, $G(f_y)$, because the $\Gk$ and $\Gl$ parameters of \Eq{maxEntSoln} will adjust to the constants $\Gd$ and $\Gf$ so that the distribution remains unchanged.

\subsection*{Intuitive aspects of invariance and measurement}

Intuitively, one can think of information invariance and measurement scale in the following way.  On a linear scale, each incremental change of fixed length yields the same amount of information or surprise independently of magnitude.  Thus, if we change the scale by multiplying all magnitudes by a constant, we obtain the same pattern of information relative to magnitude.  In other words, the linear scale is invariant to multiplication by a constant factor so that, within the framework of maximum entropy subject to constraint, we get the same information about probability distributions from an observation $y$ or $G(y)=cy$.

On a logarithmic scale, each incremental change in proportion to the current magnitude yields the same amount of information or surprise.  Information is scale dependent.  We obtain the same information at any point on the scale by comparing ratios. For example, we gain the same information from the increment $\dd y/y=\dd\log(y)$ independently of the magnitude of $y$.  Thus, we achieve information invariance with respect to ratios by measuring increments on a logarithmic scale.  Within the framework of maximum entropy subject to constraint, we get the same information about probability distributions from an observation $y$ or $G(y)=y^c$, corresponding to informationally equivalent measurements $T(y)=\log(y)$ and $T(y^c)=c\log(y)$ \cite<see>{frank10measurement}.

Nearly all of the common probability distributions arise from a simple form of information invariance, measurement scale, and a constraint on measured values of $y$ for mean values or measured values of $y^2$ for variances \cite{frank10measurement,frank11a-simple}.  The main measurement scales in the common distributions express a change of information with magnitude.  For example, the linear-log scale grades from linear at small magnitudes to logarithmic at large magnitudes, expressed as $T(y) = \log(1+by)$, where $b$ is a parameter that determines the magnitude at which the scale changes from linear to logarithmic.  When $by$ is small, the scaling is linear, and when $by$ is large, the scaling is logarithmic.  By contrast, the log-linear scale grades from logarithmic at small magnitudes to linear at large magnitudes, expressed as $T(y)=y+b\log(y)$.  

\subsection*{Gamma distribution from log-linear scaling}

When we use the log-linear scale, $T(y)=y+b\log(y)$, in the general maximum entropy solution of \Eq{shannonSoln}, we obtain the gamma distribution
\begin{align*}
  p_y &= ke^{-\Gl T_f}\\
      &= ke^{-\Gl(y+b\log(y))}\\
      &= ky^{\Ga-1}e^{-\Gl y},
\end{align*}
where $\Ga - 1 =-\Gl b$, and $k$ is the proportionality constant needed for the total probability to be one. This distribution matches the discrete gamma first presented in \Eq{discreteGammaE}. We obtain the same expression of the distribution from maximum entropy for discrete and continuous cases, with proper definition of the domain as $y=1,2,\ldots$ for the discrete case and $y>0$ for the continuous case, and using the proper normalization for $k$ to guarantee that the total probability is one in each case.

The constraint associated with log-linear scaling is $\bar{T}=\hat{\Gm} + b\log(\hat{\Gg})$, where $\hat{\Gm}$ is the mean of $y$, and $\hat{\Gg}$ is the geometric mean for which $\log(\hat{\Gg})$ is the mean of $\log(y)$.  The means may either be estimated from a sample or assumed a priori to take on particular values.  It is well known in the maximum entropy literature that one can derive a gamma distribution by constraining the mean and geometric mean \cite{kapur89maximum-entropy,frank09the-common}.  However, the measurement theory approach to maximum entropy derives the joint constraint on the mean and geometric mean as an outcome of a general method to analyze the relation between information and magnitude \cite{frank10measurement,frank11a-simple}. The earlier studies simply invoked those constraints as a sufficient description of the gamma distribution \cite{kapur89maximum-entropy,frank09the-common}. 

The form of a probability distribution under maximum entropy can be read directly as an expression of how the measurement scale changes with magnitude \cite{frank11a-simple}.  From the general solution in \Eq{shannonSoln}, linear scales $T(y)\propto y$ yield distributions that are exponential in $y$, whereas logarithmic scales $T(y)\propto c\log(y)$ yield distributions that are linear in $y^c$.  Exponential distributions of the form $e^{-\Gl y}$ arise from underlying linear scales, whereas power law distributions of the form $y^{-c}$ arise from underlying logarithmic scales.  

The gamma distribution has form $y^{-c}e^{-\Gl y}$.  When the magnitude of $y$ is small, the shape of the distribution is dominated by the power law component, $y^{-c}$.  As the magnitude of $y$ increases, the shape of the distribution is dominated by the exponential component, $e^{-\Gl y}$.  Thus, the underlying measurement scale grades from logarithmic at small magnitudes to linear at large magnitudes.  Indeed, the gamma distribution is exactly the expression of an underlying measurement scale that grades from logarithmic to linear as magnitude increases.

We can now state the key observation about measurement and ecological pattern.  Empirically, the evidence strongly demonstrates that SADs almost always follow an approximately log-linear scaling.  Variation in the transition between the logarithmic and linear regime describes nearly all of the variation in observed pattern.    

\section*{Different ecological mechanisms lead to log-linear scaling}

What are the set of underlying ecological mechanisms and aspects of measurement that lead in the aggregate to log-linear scaling of SADs?

My main goal for this paper is to reformulate the problem of SADs in terms of this question about log-linear scaling.  I cannot answer this question at present, because there is no general understanding of the different kinds of processes that, in the aggregate, lead to particular information invariances and measurement scales.  Future progress in understanding biological pattern depends strongly on progress in understanding aggregation, invariance, and scale. 

The following section presents a preliminary example. That example hints at the range of processes leading to log-linear scaling.

\section*{Stochastic models of population growth}

\citeA{dennis84the-gamma} showed that stochastic fluctuations of population growth often lead to a gamma distribution of population abundance. Their work generalized mathematical results from previous studies \shortcite{may74stability,may78exploiting}. \citeA{costantino81gamma} supplemented their mathematical derivation of gamma population abundance with supporting data from several laboratory studies of the flour beetle.  

In this section, I highlight the essential aspect of population dynamics that leads to the gamma distribution.  The essence reduces to log-linear scaling of perturbations, supporting my claim that a clear understanding of the proper scale of measurement leads to a clear understanding of biological pattern.  Measurement also helps to explain why some patterns are so common, because the most common patterns associate with the most common measurement scales.  

By studying the example of population growth, we can learn how aspects of dynamics associate with scale.  However, one must remember that this particular model of population growth is just one example of a process that leads to log-linear scaling and the gamma pattern. I give a simplified version of \citeA{dennis84the-gamma}, adding my own interpretation with regard to log-linear scaling.  

A deterministic model of population growth can often be expressed as
\begin{equation*}
  \frac{\dd y}{\dd t} = yg(y),
\end{equation*}
where $y$ is population size, and $g(y)$ is the growth rate of the population as a function of its size.  An equilibrium occurs when $g(y^*) = 0$, and the equilibrium is stable when $g'(y^*) < 0$. A model with stochastic fluctuations can be written as
\begin{equation*}
  \frac{\dd y}{\dd t} = yg(y)+yz(t),
\end{equation*}
where $z(t)$ is a Gaussian perturbation with mean zero and variance $\Gs^2$.  The magnitude of the perturbation scales linearly with population size, which is equivalent to a constant magnitude of perturbation per individual independent of population size.

Suppose we approximate the growth rate by a linear expression, $g(y) \approx a -cy$.  This linear approximation gives the standard logistic growth equation, usually with notation $a=r$ and $c=r/K$, where $-r/K$ is the slope of the growth rate as the population size increases, and $K=y^*$ is the equilibrium.  It is common to call $r$ the intrinsic rate of increase and $K$ the carrying capacity, although here those parameters simply arise from a linearization of the general function for growth rate, $g(y)$.  Putting the pieces together, we obtain a model with Gaussian stochastic fluctuations in proportion to population size and a linear approximation of population growth 
\begin{equation*}
  \frac{\dd y}{\dd t} = y(a-cy) + yz(t).
\end{equation*}
\citeA{turelli77random} analyzed various interpretations of this stochastic equation.  For my purposes, the following simplified presentation captures the essential aspect of implicit log-linear measurement scale that leads to a gamma distribution of population sizes.

We can think of perturbations of population size at each point in time as having two components.  First, a direction component arises from the tendency for the population to follow the expected instantaneous growth rate $a-cy$, leading to a mean directional tendency on population size of $m(y)=y(a-cy)$.  The stochastic perturbations, $z(t)$, have a mean value of zero and do not contribute to a directional tendency.  Second, the term $z(t)$ contributes a directionally unbiased fluctuation with variance $\Gs^2$, and the combination $yz(t)$ contributes variance $v(y)=y^2\Gs^2$. The stochastic fluctuations, $\Gs^2$, are of the same order of magnitude or less than the maximal directional tendency, $a$. 

Because the population has a randomly fluctuating component, the population size never settles to a single value.  Instead, a steady state may be reached, such that for each population size, $y$, the tendency of the population to move to another size is balanced by the overall tendency of other population sizes to change to $y$.  This balance allows us to calculate the steady state probability distribution for population sizes, which is
\begin{equation}\label{eq:stoch1}
  p_y = ke^{-\Gp(y)},
\end{equation}
where $k$ is chosen so that the total probability is one. From a general solution of stochastic differential equations \cite{may74stability,may78exploiting}, we have
\begin{equation}\label{eq:MaySoln}
  -\Gp(y) = -\log v(y) + 2\int^y\frac{m(\tilde{y})}{v(\tilde{y})}\,\dd \tilde{y}.
\end{equation}
Using the expressions above for the mean, $m$, and variance, $v$, of perturbations, and carrying out the integration, we obtain
\begin{equation}\label{eq:psiSoln}
	-\Gp(y) = (\Ga-1)\log y -\Gl y + C,
\end{equation}
where $\Gl=2c/\Gs^2$ and $\Ga-1=2(a/\Gs^2-1)$, and $C$ is a constant that will be absorbed by $k$ in the following step.
Using these expressions in \Eq{stoch1}, we obtain
{\setlength\arraycolsep{0.1em}
\begin{eqnarray*}
  p_y &=& ke^{(\Ga-1)\log y-\Gl y}\\
      &=& ky^{\Ga-1}e^{-\Gl y},
\end{eqnarray*}}
which is a gamma distribution.

My main argument is that, to interpret a probability distribution with respect to underlying process, we must focus on the measurement scale expressed by $T(y)$.  By comparing \Eq{stoch1} for the stochastic solution of the probability distribution with \Eq{shannonSoln} for the maximum entropy probability distribution in relation to measurement scale, we have that $\Gp(y)=\Gl T(y)$, where in \Eq{shannonSoln} I use $f(y)=y$ and $T_f=T(y)$.  From \Eq{MaySoln}, moving $\log v(y)$ inside the integral yields
{\setlength\arraycolsep{0.1em}
\begin{eqnarray*}
 -\Gl T(y) = -\Gp(y) = 2\int^y\frac{m(\tilde{y})-v'(\tilde{y})/2}{v(\tilde{y})}\,\dd \tilde{y},
\end{eqnarray*}}
where $v'$ is the derivative with respect to $y$.  Differentiating and using the shorthand notation $m\equiv m(y)$ and $v\equiv v(y)$, we obtain
\begin{equation*}
  \dd T(y) \propto \frac{m-v'/2}{v}\,\dd y.
\end{equation*}
The directional and stochastic perturbations, $m$ and $v$, change the effective measurement scale in the manner given by this expression. One may think of the scale $\dd T$ as expressing information in relation to magnitude.  Stronger stochastic fluctuations, $v$, effectively reduce the precision or information with regard to directional tendency, whereas stronger directional fluctuations, $m$, in relation to $v$, effectively provide more information about directional tendency. 

Returning to the specific problem of population growth that leads to the gamma distribution, we can translate the stochastic solution for the probability distribution based on \Eq{psiSoln} into an expression in terms of the measurement function $T(y)$, yielding
\begin{equation*}
	-\Gl T(y) = -\Gp(y) = (\Ga-1)\log y -\Gl y,
\end{equation*}
allowing us to write the measurement function in the generic form of log-linear scaling, 
\begin{equation}\label{eq:logisticMeasure}
  T(y) = y + b\log y,
\end{equation}
or, equivalently,
\begin{equation*}
  \dd T(y) \propto \left(1 + \frac{b}{y}\right)\dd y.
\end{equation*}
This measure scales logarithmically when abundance, $y$, is low and linearly when abundance is high.  

The parameter $b=-(\Ga-1)/\Gl$ determines the magnitude of abundance at which the scale grades from logarithmic to linear.  We can express $b$ in terms of the parameters of the growth equation and the magnitude of stochastic perturbations 
\begin{equation*}
  b = -K\left(1-\frac{\Gs^2}{r}\right),
\end{equation*}  
where I have used the traditional parameters of logistic growth: $r$ for intrinsic rate of increase, and $K$ for carrying capacity (see above).  

This expression for $b$ shows that higher carrying capacity, $K$, increases the domain of the logarithmic regime to higher magnitude, causing pattern to follow a power law over a wider span.  This control of the relative logarithmic and linear domains is easy to understand.  At low abundance relative to carrying capacity, the deterministic component of population growth perturbs abundance exponentially.  As abundance approaches the carrying capacity, deterministic perturbations become linear in abundance. 

The term $\Gs^2/r$ reflects the relative scaling of the linear stochastic perturbations, $\Gs^2$, to the logarithmic deterministic perturbations, $r$.  Greater relative stochastic fluctuations reduce $b$ and shift scaling toward the linear domain, because noise in this model is added as a linear perturbation.

\section*{Discussion}

The gamma distribution expressed in terms of the log-linear measurement scale can be viewed in two alternative ways.  First, the derivation of the gamma distribution from a model of stochastic population growth can be thought of as the primary line of reasoning.  By that view, the log-linear interpretation follows secondarily as a description of stochastic population growth.  Second, one may think of the log-linear measurement scaling as a primary argument for why a gamma distribution is likely to be a common pattern.  By that view, the derivation from a particular stochastic model of population growth arises secondarily as a special instance of a much wider class of problems that share the common log-linear scaling.  

The current literature promotes the first view: primary, specific derivations from underlying mechanistic models.  In my opinion, that approach is certain to be exactly wrong.  The exact part arises because the models derive exact expectations from explicit mechanistic assumptions.  The wrong part arises because nature will certainly not be the outcome of exactly those specific assumptions.  Natural pattern will almost certainly be dominated by the aggregation of various distinct processes.  

In thinking about such aggregation, we will never be able to specify exactly the various ecological mechanisms and their relative contributions.  We can only vaguely analyze how such aggregation may consistently shape pattern.  However, we know that aggregation of heterogeneous processes can sometimes attract strongly to particular outcomes, as in the central limit theorem.  My conjecture is that species abundance is dominated in the aggregate by a log-linear measurement scale, reflecting relatively consistent informational invariances in relation to magnitude.  Such invariance leads to the gamma pattern.  By this route of analysis, one can be vaguely but consistently right about pattern and its causes.

The ecological and genetical literature has devoted little effort to search for general principles such as measurement invariance. Instead, we have a vast catalog of specific assumptions leading to exactly specified outcomes.  The neutral theories are typically posed as outcomes that follow exactly from precise assumptions about process, rather than expectations that follow vaguely from general principles of aggregation and measurement.  These fields will never mature fully until they develop a clearer sense of the relations between theory and pattern.

My emphasis on aggregation's tendency to attract to particular patterns is of course not new.  Many authors, such as \citeA{may75patterns} and \citeA{levin97theories}, have advocated the perspective of the central limit theorem or mean field approximations to explain consistent pattern. More recently, several papers have discussed how different ecological mechanisms attract to the same species abundance distributions \shortcite<summarized by>{alonso08the-implicit}.  

However, these studies, on how different mechanisms attract to the same outcome, are limited in scope.  Such studies rarely develop a clear sense of what kinds of processes do attract to a particular outcome, such as a gamma distribution, and what kinds of processes do not attract to that particular pattern.  In this regard, measurement theory sets the important questions. What essential aspects of ecological mechanisms and aggregation preserve key information invariances?  What aspects of process do not matter, because the details of those processes do not alter the fundamental information invariances?  

The words \textit{invariance\/} and \textit{symmetry\/} can be used interchangeably, in the sense stated by \citeA[p.~4, attributed to Hermann Weyl]{cantwell02introduction}:  \textit{An object is symmetrical if one can subject it to a certain operation and it appears exactly the same after the operation.  The object is then said to be invariant with respect to the given operation.}

Thus, the causes of pattern reduce to the fundamental symmetries of ecological process.  Here, the symmetries may often have to do with scale in relation to dynamics.  For example, in the log-linear scaling that arises in the particular model of stochastic population growth, the invariant transformations that define the symmetries of the measurement scale change with magnitude, because relative scaling of the deterministic and stochastic perturbations change with magnitude.  Many problems must reduce to such descriptions of symmetries, capturing the essence of process in relation to pattern.  What we need is more work devoted to the special aspects of ecological and genetic processes, so that we can readily reduce particular, complex situations to their essential symmetries.  That view, although relatively unused in biology, is not so far-fetched.  As \citeA{anderson72more} noted:  \textit{It is only slightly overstating the case to say that physics is the study of symmetry. }

\section*{Acknowledgments}

My research is supported by National Science Foundation grant EF-0822399,
National Institute of General Medical Sciences MIDAS Program grant
U01-GM-76499, and a grant from the James S.~McDonnell Foundation.
  
\vfill\eject

\bibliography{entropy}
\bibliographystyle{apaciteJEB}

\section*{Appendix}

\citeA{pueyo07the-maximum} used maximum entropy to derive the log series distribution for species abundances.  Their maximum entropy method applied Jaynes' \citeyear{jaynes03probability} concepts of invariance and symmetry to derive the log series pattern.  From the log series, \citeA{pueyo07the-maximum} added an additional constraint to obtain the gamma distribution.  They showed that the gamma subsumes the log series, geometric, and power law distributions and is close to the lognormal.  These points are the same ones that I have emphasized throughout my paper.  In this regard, \citeA{pueyo07the-maximum} deserve full credit for the origins of these ideas and their application to species abundances.  Given this clear precedent, my presentation in this paper extends the topic in two ways.

First, \citeA{pueyo07the-maximum} begin with the same concepts of invariance and symmetry from \citeA{jaynes03probability} on which I based my own approach.  The Jaynesian approach is itself a slight extension of Jeffreys' \citeyear{jeffreys57scientific} Bayesian notion of prior distributions, in which one maximizes a measure of entropy that is taken relative to a prior description of what is most random for a given problem.  

\citeA{pueyo07the-maximum} emphasize that their particular explanation of the prior distribution for ecological problems is their most important contribution.  On page 1023, they say
\begin{quote}
\textit{We have shown that common shapes of SADs can be predicted from extremely general assumptions. This conclusion is extensive to common shapes of SARs [species area relations], because these shapes are mathematically related to the SADs we found \cite{pueyo06self-similarity}. We expect more findings to follow, because we think we have correctly identified the prior distribution (eqn 7), which is the Rosetta Stone that allows translating concepts between statistical physics and macroecology.}
\end{quote}

The way in which Pueyo \textit{et al.}\ developed the prior following Jaynes was indeed the state of the art in 2007.  However, the particular derivation they gave and their particular result was just a long way of arriving exactly at the famous Jeffreys prior, in which the ``most random'' expression of a variable, $n$, is a probability distribution proportional to $n^{-1}$.  

\citeA{pueyo07the-maximum} present various specific derivations of the Jeffreys prior based on particular ecological assumptions.  In one example that they highlight, spatial and geometric aspects play a role in their derivation.  Although they emphasize that the Jeffreys prior is a very general expression of invariance and symmetry, at the same time, they tie their expression of their key result to rather specific ecological descriptions.  

I followed the same broad concepts, but the limitations in the approach of \citeA{pueyo07the-maximum} are important.  The Jeffreys approach is based on a Bayesian notion of relative entropy with a prior notion of randomness.  Our work has shown that view to be too limiting in understanding probability \cite{frank10measurement,frank11a-simple}.  In the particular case of the Jeffreys prior, all that is involved is the assumption of ratio invariance for measurements, leading to the natural scale for information as logarithmic.  In this regard, \citeA{pueyo07the-maximum} are too specific in claiming the association between their particular ecological motivations and logarithmic scaling, giving the false impression that they have found the ``Rosetta Stone'' for ecological pattern, when in fact all that they have done is give some specific examples in which information is properly measured on a logarithmic scale, but without a clear notion of the general role of measurement.  

The understanding of ecological pattern will remain confused as long as one conflates the general issue of measurement scale with the specifics of certain ecological examples.  That confusion will prevent the full conceptual power of maximum entropy and measurement from being appreciated in its application to ecology. In  this paper, I worked toward joining the principles of measurement theory to ecological pattern.

My second extension to \citeA{pueyo07the-maximum} concerns how one can derive the gamma distribution, the general expression of SAD pattern.  \citeA{pueyo07the-maximum} first derived, too specifically, the Jeffreys prior.  From that prior, and a constraint on mean abundance, they obtain the log series.  They then recognize that the log series is too limited to describe SAD pattern.  To extend to the gamma, they note that the log series constrains the geometric mean when given arithmetic mean abundance.  To get the gamma from the log series, they realize that the geometric mean must be allowed to vary independently of the arithmetic mean, so, ad hoc, they allow the geometric mean to vary independently.  

Once the arithmetic and geometric means can vary independently, one has the well known description of the gamma distribution as arising when arithmetic and geometric means are sufficient statistics within a context of maximum entropy subject to constraint \cite{kapur89maximum-entropy}.  Although a correct expression of the relation between constraint and pattern, that ends up just being a description of the gamma rather than a derivation of the pattern from a principled way of understanding how pattern arises.  By that approach, any probability distribution can be obtained by tautologically invoking the sufficient statistics that are the constraints under maximum entropy.  One loses any claim to deriving pattern from fundamental principles.

By contrast, Eric Smith and I \cite{frank10measurement,frank11a-simple} have replaced the limited Jeffreys-Jaynes approach to relative entropy priors with a principled notion of measurement scale based on the fundamental concepts of measurement theory \cite{hand04measurement}.  We started with the standard principles of measurement theory and then developed a novel extension of measurement to show the structural relations between a variety of common measurement scales.  Those structurally related measurement scales combined with maximum entropy encompass essentially all of the common families of probability distributions.  By contrast, \citeA{pueyo07the-maximum}, by following the standard implementation of the Jeffreys-Jaynes approach, found that their log series expression had to be supplemented with an ad hoc assumption to get to the desired gamma pattern. In that one arbitrary step, they are already outside of a coherent framing of invariance, symmetry, and pattern.  

All of this may seem more important for probability and mathematics than for ecology.  However, understanding ecological pattern depends on understanding how aggregation, measurement, and randomness set the basic contours of pattern in nature.  In this regard, the recent influx of maximum entropy concepts into ecology is a welcome step.  But if those concepts are developed in ecology in a limited way or, more commonly, in a way that confuses the specific and the general, the net result will be the tendency to reject maximum entropy as a failed or confused approach.  That would be a mistake, because ecological understanding will necessarily be limited if the field does not properly incorporate the fundamental principles of measurement and information.

\end{document}